\documentclass[10pt,journal,compsoc]{IEEEtran}
\usepackage{xcolor}
\usepackage{paralist}
\usepackage{url}
\usepackage[T1]{fontenc}
\usepackage[utf8]{inputenc}
\usepackage{booktabs}  
\usepackage{xcolor}
\usepackage{tcolorbox}
\usepackage{xifthen}
\usepackage{enumitem}
\usepackage{microtype}
\usepackage{graphicx}
\usepackage{adjustbox}
\usepackage{makecell}
\usepackage{textcomp} 
\usepackage{pifont} 
\usepackage{transparent}

\everypar{\looseness=-1}

\newcommand{\QUOTE}[1]{``\textit{#1}''}

\newcommand{\myparagraph}[1]{\noindent\textbf{#1}}
\newcommand{\INTERPRETATION}[1]{\noindent\textcolor{black}{\fcolorbox{gainsboro}{gainsboro}{\begin{minipage}[t]{.97\columnwidth}\textbf{Interpretation: }#1\end{minipage}}}}

\newcommand{\NRPapers}{13}
\newcommand{\NRParticipants}{18}
\newcommand{\NRDesignDimensions}{seven}
\newcommand{\NRDesignQualities}{ten}
\newcommand{\concept}{criteria}

\renewcommand{\arraystretch}{1}

\definecolor{pastelgray}{rgb}{0.81, 0.81, 0.77}
\definecolor{beaublue}{rgb}{0.74, 0.83, 0.9}
\definecolor{mossgreen}{rgb}{0.88, 0.97, 0.88}
\definecolor{whitesmoke}{rgb}{0.96, 0.96, 0.96}
\definecolor{pastelgray}{rgb}{0.81, 0.81, 0.77}
\definecolor{beaublue}{rgb}{0.74, 0.83, 0.9}
\definecolor{gainsboro}{rgb}{0.86, 0.86, 0.86}
\definecolor{cadetgrey}{rgb}{0.57, 0.64, 0.69}
\definecolor{carolinablue}{rgb}{0.6, 0.73, 0.89}

\newcommand*{\myuparrowt}{%
	\includegraphics[
	height=.9\baselineskip,
	width=.9\baselineskip,
	keepaspectratio,
	]{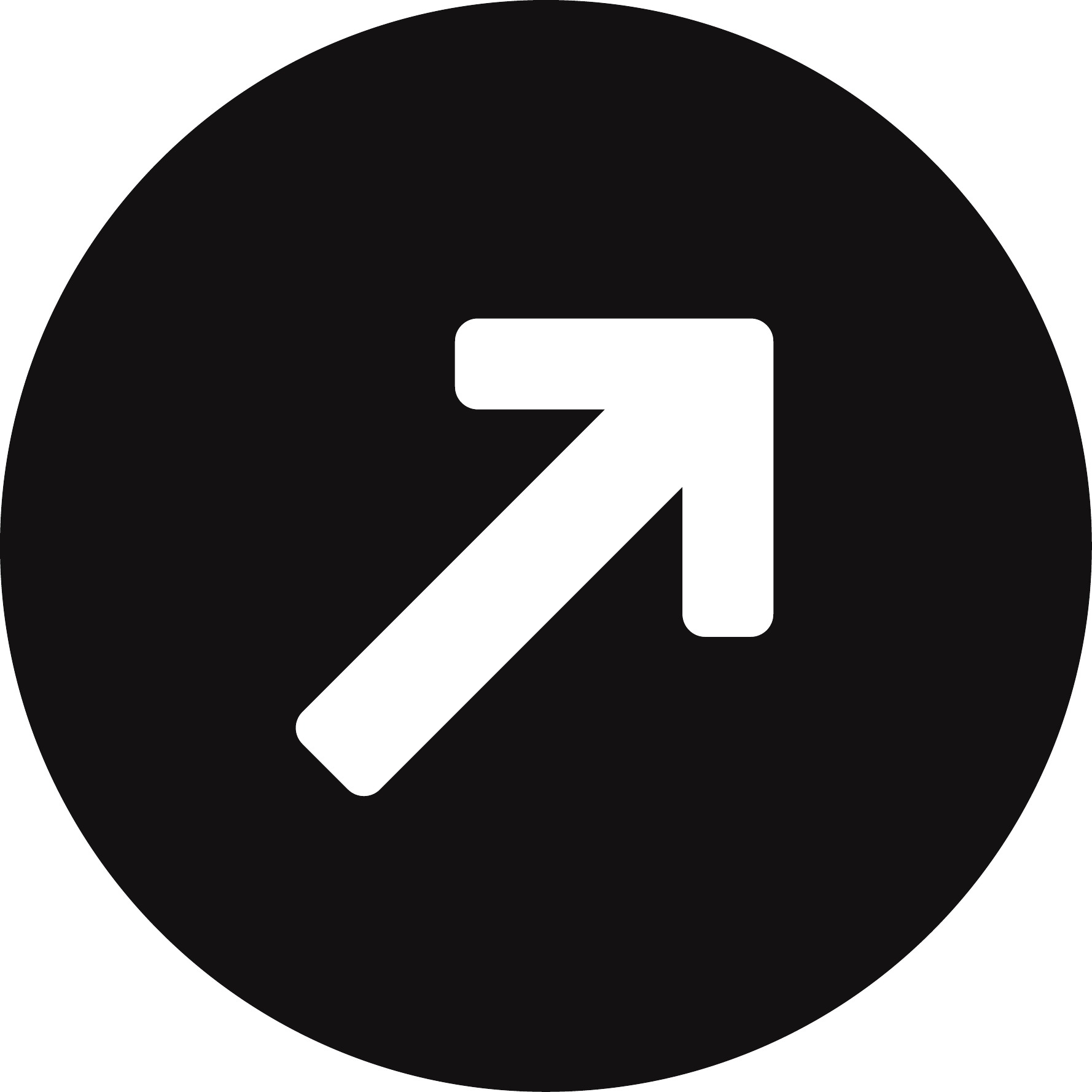}%
}

\newcommand*{\mydownarrowt}{%
	\includegraphics[
	height=.9\baselineskip,
	width=.9\baselineskip,
	keepaspectratio,
	]{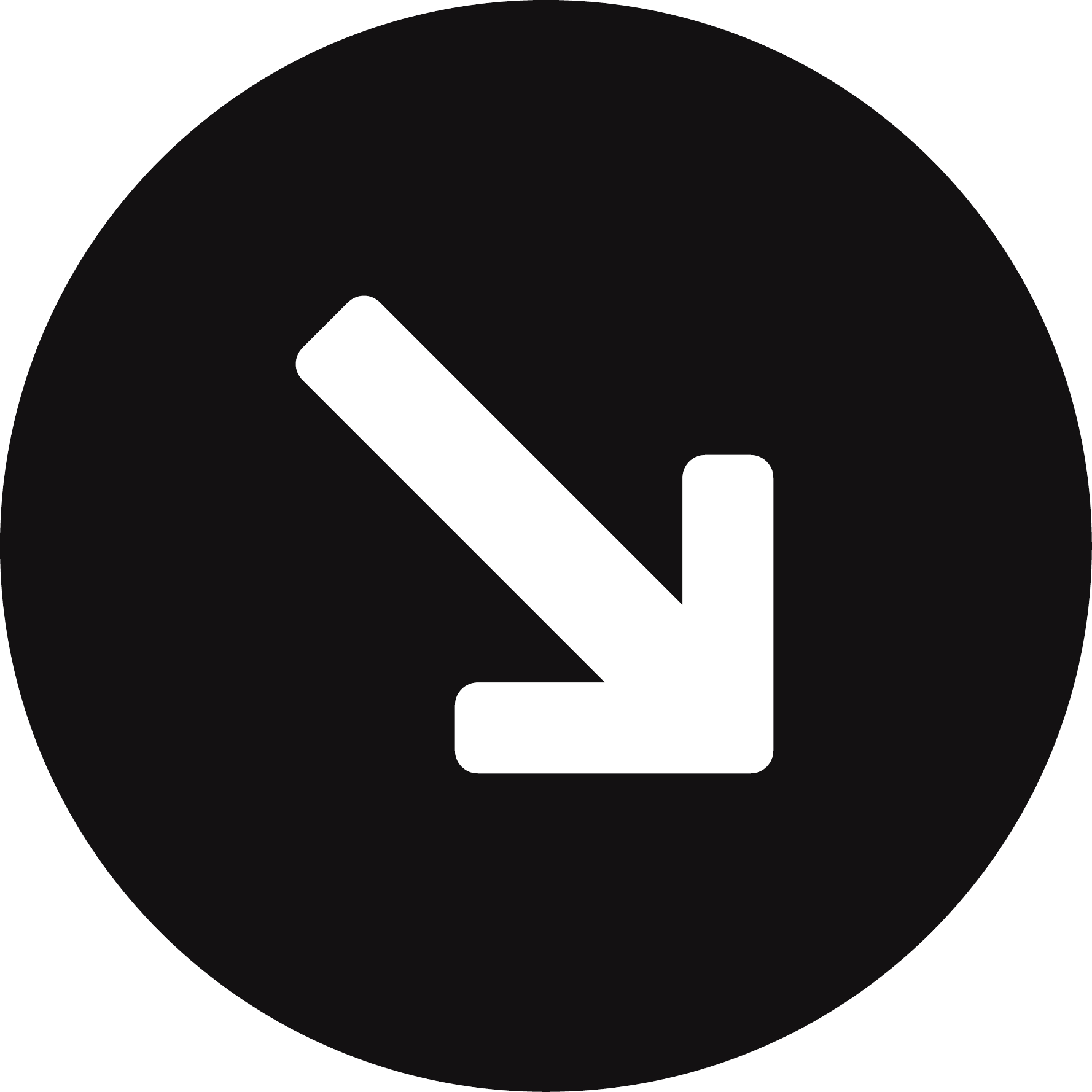}%
}

\newcommand*{\myquestionmarkt}{%
	\raisebox{-.05\baselineskip}{%
		\includegraphics[
		height=.92\baselineskip,
		width=.92\baselineskip,
		keepaspectratio,
		]{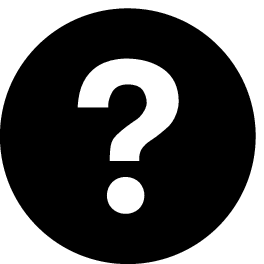}%
	}%
}

\newcommand*{\myuparrow}{%
	\raisebox{-.2\baselineskip}{%
		\includegraphics[
		height=.8\baselineskip,
		width=.8\baselineskip,
		keepaspectratio,
		]{figures/right-up}%
	}%
}

\newcommand*{\mydownarrow}{%
	\raisebox{-.2\baselineskip}{%
		\includegraphics[
		height=.8\baselineskip,
		width=.8\baselineskip,
		keepaspectratio,
		]{figures/right-down}%
	}%
}

\newcommand*{\myquestionmark}{%
	\raisebox{-.1\baselineskip}{%
		\includegraphics[
		height=.8\baselineskip,
		width=.8\baselineskip,
		keepaspectratio,
		]{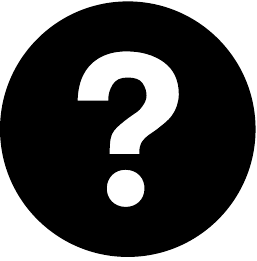}%
	}%
}

\begin{document}
%
\title{Design Dimensions for Software Certification:\\ A Grounded Analysis}

\author{Gabriel~Ferreira, Christian~K{\"a}stner, Joshua~Sunshine, Sven~Apel,~and~William~Scherlis}


\IEEEtitleabstractindextext{%
\begin{abstract}
In many domains, software systems cannot be deployed until authorities judge them fit for use in an intended operating environment. Certification standards and processes have been devised and deployed to regulate operations of software systems and prevent their failures. However, practitioners are often unsatisfied with the efficiency and value proposition of certification efforts. In this study, we compare two certification standards, Common Criteria and DO-178C, and collect insights from literature and from interviews with subject-matter experts to identify design options relevant to the design of standards. The results of the comparison of certification efforts---leading to the identification of design dimensions that affect their quality---serve as a framework to guide the comparison, creation, and revision of certification standards and processes. This paper puts software engineering research in context and discusses key issues around process and quality assurance and includes observations from industry about relevant topics such as recertification, timely evaluations, but also technical discussions around model-driven approaches and formal methods. Our initial characterization of the design space of certification efforts can be used to inform technical discussions and to influence the directions of new or existing certification efforts. Practitioners, technical commissions, and government can directly benefit from our analytical framework.
\end{abstract}

\begin{IEEEkeywords}
software certification, certification standards design, interviews, grounded analysis
\end{IEEEkeywords}}

\maketitle
\IEEEdisplaynontitleabstractindextext
\IEEEpeerreviewmaketitle

\IEEEraisesectionheading{\section{Introduction}\label{sec:introduction}}

\sloppy
Software certification by regulatory entities plays a fundamental role in defining and assessing the fitness of software systems in environments where failures can result in serious consequences. Software flaws can negatively affect business, environment, and society, for example, when confidential data are leaked or when medical devices emit excessive radiation \cite{therac25:1993, dowson:1997, risksToThePublic:2016}. 

Airborne and power plant systems have a long regulatory history, and their requirements and development processes are better understood 
than those from emerging domains such as software-based medical devices, IoT ecosystems, or self-driving cars. New guidelines are being discussed to address the challenges posed by these domains, but these are still incipient initiatives. There is much discussion but little agreement on how these domains should be regulated. There are many competing drivers for regulating a domain, defining standards, and establishing certification processes for software systems. For example, different stakeholders expect certification processes to:
\begin{itemize}
	\item enhance the quality of products,
	\item specify software processes to guide development of products,
	\item guide requirements definition for certified products and define a common vocabulary among practitioners, and
	\item serve as a legal safe harbor.
\end{itemize}

At the same time, current certification schemes are often criticized for creating perverse incentives \cite{anderson:2001, hearn:2004, thomas:2007, dodd:2012}. For example, Common Criteria allows vendors to select testing laboratories; vendors may then select laboratories that give ``easy'' evaluations, by charging less, being less thorough, or taking less time; contradicting the ``enhance the quality of software products'' goal of the certification efforts \cite{anderson:2001}. The value proposition of certification is also questioned. For example, even experts claim that \QUOTE{belief in the effectiveness of DO-178C and similar standards is superstition, not science or engineering} \cite{thomas:2007}. As a consequence, certification is sometimes seen as just a ``tick the box'' process and as an obstacle for selling products \cite{hearn:2004}. In cases where vendors are satisfied with the value proposition of certification efforts, there are still efficiency and scalability issues. For example, the long development cycles and certification time observed in the aerospace industry might be incompatible with the automobile industry's expectations \cite{nhtsaFederalAutomatedVehicles}. 

Another common criticism of current certification approaches is that certification methods can be subjective and do not make use of the most powerful or cost-effective quality assurance methods to produce or evaluate evidence. In our study, we use \emph{evidence} to refer to any artifact that is produced for quality-evaluation purposes. Software certification often focuses on evidence derived from documenting and manually analyzing requirements, development processes, and test protocols. Advances in software engineering techniques, for example, testing, theorem proving, and static analysis, which could mechanize evidence production and provide more guarantees, are only slowly adopted, if at all.
Many argue that the next generation of certification standards must focus on 
making the use of such techniques ubiquitous to reduce subjective 
interpretation and increase objective production of evidence 
\cite{swForDependableSystemsBook, maibaum:2008, hatcliff:2009}. 

Despite all the tension around existing and upcoming certification standards, 
there is little public knowledge about what aspects of designing certification efforts 
are important or controversial. Domain experts are often
familiar with a single standard and its evolution and many
have published white papers and magazine 
articles~\cite{lipner:1991, hearn:2004, keblawi:2006, thomas:2007, maibaum:2008, rushby:2011, kallberg:2012}, 
often calling for change to a specific standard. However, domain 
experts have their own agenda and might not be aware of alternatives adopted in 
other domains or have a limited view of the conflicting forces at play. 

Our goal is to step back and to compare two certification efforts based on the analysis of standard documents and on the perceptions of stakeholders involved with them in their everyday activities. Specifically, we performed a multiple case study \cite{yin:2009caseStudyBook} based on two certification standards: \emph{Common Criteria}, which assesses security requirements of infrastructure and
end-user software and devices, and \emph{DO-178C}, which assesses safety requirements
in software aerospace systems. We specifically chose standards from two distinct communities to elicit alternatives adopted by them and to increase stakeholders awareness of the conflicting forces at play on each community.

For this purpose, we cross-validated data from three sources: (i) interviews with \NRParticipants~ experts working with these standards in various roles, (ii) analysis of standards' documents, and (iii) a survey of 11~ papers discussing these standards (often white papers and magazine articles).

Our results contribute to an \emph{analytical framework} grounded in data that can support the creation and revision of certification efforts. The framework elicits relevant design dimensions, their range of options, and the interactions that arise between options and the quality of certification efforts. We also characterize each standard in the framework as we identify their alternatives in the design dimensions, but also highlight points of contention, trends, and open questions for designers of future certification efforts. Our results are intended to better-inform decisions about software processes, techniques, and tools that affect certification efforts design.

Our paper does not intend to present or evaluate solutions for the design of certification standards (the evaluation on a new standard might be a 10-year agenda beyond the scope of a single paper), but it encourages further investigation and ground discussions with data from two important cases \cite[ch.~13]{brooks:2010}, which is a substantial step toward a testable theory and valuable in itself for the community. 


In summary, we contribute 
\begin{inparaenum}[(1)]
	\item insights from interviews with \NRParticipants~ experts and cross-validated with 
	analysis of the standards' documents and a literature survey,
	\item an analytical framework with design 
	dimensions and interactions between them for creating and assessing certification efforts,
	\item an analysis of two standards in the context of the framework; 
	including design options collected from \NRParticipants~ interviews 
	as well as from \NRPapers~ papers written by subject-matter 
	experts, and
	\item discussion of challenges in certification efforts and implications 
	for future standards.
\end{inparaenum}

\section{Certifying Software Systems}
\label{sec:certifyingSoftSystems}
Certification standards define the technical baseline for vendors to produce 
compliant products that demonstrate a set of desired quality attributes and for evaluators to judge their compliance to these attributes. In this 
context, software represents both the means and end of certification. Many 
certification efforts exist for various kinds of properties. Common Criteria \cite{commonCriteria:CC}, DO-178C \cite{do178c}, IEC 62304 \cite{isoIEC62304}, and NIST RMF 800 Series \cite{nistRMF800}, for example, play a leading role in 
the certification of infrastructure and application software systems, airborne 
systems, medical device software, and information security controls for systems and 
organizations, respectively. They differ in formality, adoption, and 
certification processes, with focus ranging from product requirements to life cycle 
development and maintenance requirements to organizational requirements.

Early certification standards with focus on product requirements optimistically expected that mathematical models and system design elements could enable the development of systems that were virtually free of safety or security violations \cite{thomas:2007, yost:2007}. Certification standards in the safety community were inspired by hardware standards, often associating statistical failure rates with assurance levels to address the criticality of products. In contrast, certification standards in the security community were originally motivated by the proliferation of time-sharing computing, which replaced batch processing as a paradigm and defined multilevel computer security as field \cite{yost:2007, yost:2015}. Today it is broadly accepted that it is unlikely to achieve the probability of failures level and the ``vulnerability-free'' status to which early certification efforts aspired \cite{lipner:2004, thomas:2007}.

More recent certification standards relax the ``high-assurance'' expectations of earlier ones. Instead of aspiring to guarantee the correctness of a system, they \emph{provide the means to analyze the design and implementation of functionality by checking how particular requirements are met}; that is, they focus on specific requirements. For our study, we investigate two current and broadly used standards, \emph{Common Criteria} and \emph{DO-178C}, that follow this mindset.

\subsection{Common Criteria}
\label{sec:commonCriteria}
Common Criteria \cite{commonCriteria:CC} is an international computer security standard created to evaluate security attributes of products against security specification and requirements. 
Systems certified with Common Criteria are typically infrastructure software, such as databases, operating systems, and firewalls, as well as various software systems and devices for networking, trusted computing, digital signatures, smart cards, and biometric systems.
Common Criteria originated from the U.S. TCSEC standard (Orange Book), released in 1982, and was later unified with the European ITSEC standard to provide mutual recognition mechanisms to products addressing international markets \cite{kallberg:2012, lipner:2015birthAndDeathOrangeBook}. For Common Criteria, security requirements such as \emph{information security principles} and \emph{access control} are the main quality attributes of concern. Functional testing, vulnerability analysis, and secure development practices \cite{howard:2006:SDLbook,commonCriteria:CC} are the main proxies for assessing security.


A certification in Common Criteria works as follows. A document called a \emph{Security Target} defines the scope of an evaluation and describes a set of implementation-dependent security requirements.
To achieve consistent standards across multiple products in one domain, \emph{Protection Profiles} (PP) describe technology-specific and implementation-independent security requirements for many domains, such as operating systems, databases, and VPN clients. 
For example, a protection profile for Web Browsers specifies requirements for secure cookie handling, deleting browser data, and sandboxing the rendering process among many others~\cite{niap2014browser}.

The roles defined by certification with Common Criteria are:
\begin{inparaenum}[(1)]
	\item developers/vendors produce the required evidence for a product evaluation, including the \emph{Security Target},
	\item evaluators/testing laboratories are private companies that evaluate products in accordance with the policies defined by a certification scheme, and
	\item meta-evaluators are responsible for the oversight of testing laboratories and to ensure consistency among them.
\end{inparaenum}

A typical evaluation process in Common Criteria works as follows: 
\begin{inparaenum}[(1)]
	\item the vendor identifies what evidence needs to be produced for a given set of security requirements,
	\item the vendor produces a \emph{Security Target}, the evidence that the product meets it, and submits both to a testing laboratory,
	\item a testing laboratory then checks the compliance between the two and performs independent testing and vulnerability analysis of the product.		
\end{inparaenum}

The standard defines seven levels of assurance for security requirements (EAL-1 to EAL-7), which require increasingly detailed documentation and assessment methods. For example, demonstrating that the implementation of a system corresponds to its design, as the \emph{ADV\_TDS} assurance family specifies, can range from informal design description (EAL-2) to formal specification (EAL-7). \emph{Protection Profiles} created in Europe specify assurance levels \cite{commonCriteria:CC} in their description that can be selected for evaluations according to customers need. However, \emph{Protection Profiles} defined in the United States do not allow the selection of assurance levels and accurately specify the minimal set of requirements and assurance for a technology. Recently, evaluations conducted in the United States are accepted only if a product claims strict conformance with an approved Protection Profile \cite{niap2018faq}.

\subsection{DO-178C}	
\label{sec:DO178C}
DO-178C \cite{do178c} is a commercial standard used to regulate the development and certification of software-based aerospace systems. It
is currently used to evaluate airborne software systems by aviation authorities around the world. The standard is not freely available; so we joined the Radio Technical Commission for Aeronautics (RTCA) to gain access. DO-178C is a 2011 replacement of DO-178B (introduced in 1992). For DO-178C, general \emph{safety, reliability} and \emph{predictability} are the key quality attributes of concern. 
As a proxy for these qualities, \emph{traceability} between requirements and source code and \emph{deterministic resource consumption} are evaluated.

The roles defined by certification with DO-178C are:
\begin{inparaenum}[(1)]
	\item developers/vendors are usually aerospace manufacturers and suppliers; they produce the required evidence for product evaluation,
	\item evaluators/designated engineering representatives (DERs) are private companies or public agencies accredited by aviation certification authorities; they evaluate products in accordance to the standard throughout four stages of development \cite{faaJobAid}, and
	\item meta-evaluators/certification authorities oversee designated engineering representatives and their processes.
\end{inparaenum}
For uniformity, from now on we refer to ``designated engineering representatives'' as ``testing laboratories.''

A typical evaluation process in DO-178C is similar to one in Common Criteria---vendors generate evidence and evaluators check the evidence to qualify the associated system. A difference is that DO-178C defines five levels of assurance determined by safety assessment processes, from A to E with decreasing degree of criticality. These levels reflect the maximum severity of failure in the software under evaluation and, consequently, the thoroughness of the evaluation process, as measured by the number of objectives met and the degree of independence required between development and assurance team to verify the objectives. For example, passenger entertainment and internet communications subsystems, being instances of low-criticality systems, are evaluated according to level E objectives.

\section{Research Design}
\label{sec:researchDesign}
This study builds on the experience of the authors in engaging with diverse stakeholders from government and industry who are concerned with the development of confident assurance judgments. Initial discussions with stakeholders, often based in opinions rather than facts, motivated our grounded research approach based on interview data, literature survey, and certification standards document analysis.

We contrast two distinct software certification standards aiming at characterizing the design space \cite{tatar:2007designTensions, brooks:2010, shaw:2012} of certification efforts. Rather than generating a complete and bounded set of independent design options for designing certification efforts, we aim at start eliciting relevant design dimensions across two existing cases, namely Common Criteria and DO-178C. The characterization of a design space for certification efforts provides an analytical framework that can be used to inform technical discussions towards revising or designing new standards. Design space representations have long been used to codify knowledge about families of designs \cite{shaw:2012, card:1990:DSInputDevices, kern:2009:DSAutomotiveUI, romer:2004:DSWirelessComm, xie:2006:DS3DArch, brun:2013:DSSelfAdaptSys}.

Our investigation focuses on the comparison of two certification standard cases, namely Common Criteria and DO-178C, and is centered around two research questions:

\noindent\begin{compactitem}
	\item \textbf{RQ1:} What are the design dimensions for certification efforts?
	\item \textbf{RQ2:} How do design dimensions affect the quality of certification efforts?
\end{compactitem}

\looseness=-1
To answer these two questions and to understand how design dimensions affect the efficiency and effectiveness of certification efforts, we combine qualitative interviews of domain experts with iterative data analysis. We use this method because it allows us to combine in-depth open-ended inquiry with focused data analysis. This allows insights to emerge iteratively and enables us to have more control over the data being constructed in the study than other methods \cite{charmaz:2006ConstructingGTbook}. A survey, for example, cannot provide the same opportunities for reflection and would require an upfront theory \cite{shull:2007guideAdvancedEmpiricalSEBook}. We follow standard methods of qualitative research that grounds results and discussion in data, but do not claim to follow any grounded theory model because we start our study with a preconceived set of comparison criteria.


Our paper aims at contrasting two communities/standards and at presenting cross-case insights. We gather opinions from practitioners in different roles, organizations, and countries, aiming to include diverse backgrounds and experiences. We look beyond a single standard to understand the broader context, aiming to achieve \emph{theoretical replication} \cite{yin:2009caseStudyBook}. As characterized by Yin, the goal with theoretical replication is to study important cases that can contribute to analytic generalization and provide significant insights for further research rather than achieve statistical generalization. Instead of discussing each case individually, we aim at examining differences in processes and outcomes of certification efforts across communities and at presenting \emph{cross-case} insights and conclusions about their design.

In the rest of this section, we provide a more detailed description of: 
\begin{inparaenum}[(1)]
	\item our research method based on interviews and content analysis,
	\item our participant recruitment strategy,
	\item our paper selection process for the literature survey, and
	\item threats to validity.	 	
\end{inparaenum}

\subsection{Research Method}
\label{sec:researchMethod}

\begin{figure}[t]
	\centering
	\includegraphics[width=\linewidth]{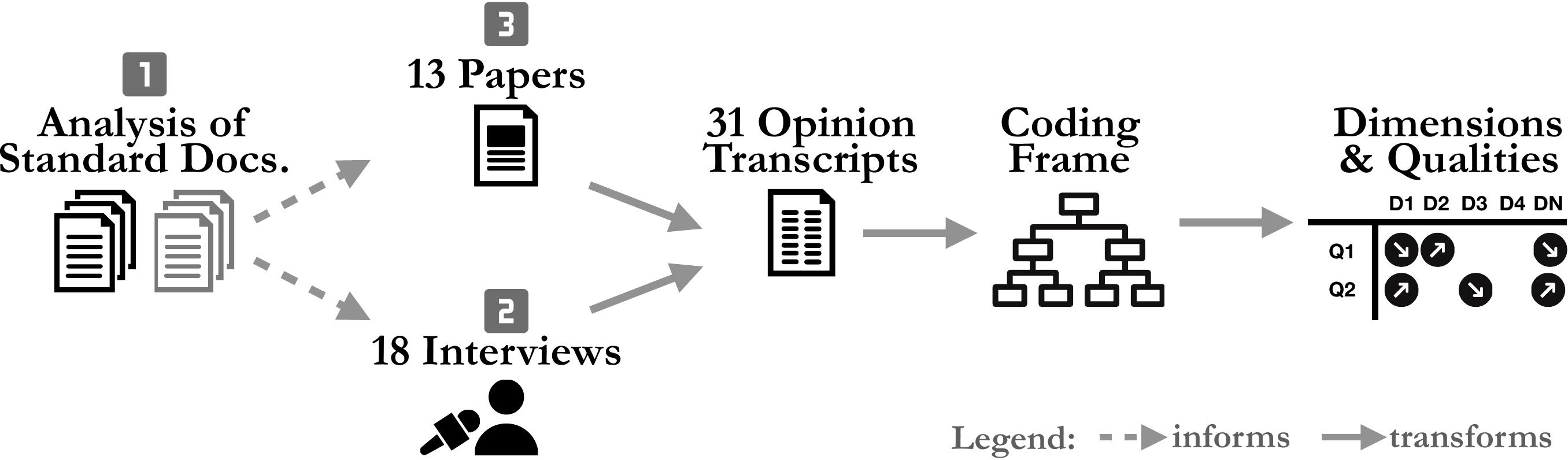}
	\caption{Overview of our research method: from the initial standard document analysis, to interviews and literature survey, to the final design dimensions and qualities.}
	\label{fig:method-overview}
\end{figure}

As is common in qualitative exploratory research, we proceeded in multiple iterations. We started conceptualizing our prior understanding from informal discussions about software certification into a list of \concept{} for further study. Next, we studied the documents of both certification standards and performed a literature survey on both standards, applying and refining the list of \concept{} (see Appendix \ref{apdx:analysisStdDocuments}). We then used these \concept{} to derive initial interview questions and conducted interviews with \NRParticipants~experts that worked with the standards in different roles (see Appendix \ref{apdx:interviewScripts}). We interleaved coding and analysis of the interview transcripts with revising our questions catalog. After concluding all interviews, we again analyzed all documents to identify common themes and coded all interview transcripts and papers. We describe the specifics of each step next. Figure \ref{fig:method-overview} shows an overview of the followed research method.

First, we collected a set of \concept{} that could be used to guide our comparison of certification efforts (see Table \ref{tab:setofCriteria}). These \concept{} represent potential points of discussion about certification efforts and were based on our experience and on prior discussions with stakeholders in the field. It was the very motivation of this study to confront ours and their beliefs and preconceptions \cite{devanbu:2016}, approaching observations around certification efforts more scientifically by grounding them on data . Our initial set of \concept{} was:

\begin{table}[t]
	\centering
	\footnotesize
	\caption{\small Initial set of \concept{} used to design our study.}
	\begin{tabular}{p{8.2cm}} \toprule 
		Quality attributes addressed; level of quality assurable. \\
		Validity of certification results.\\
		Reliability of evaluations and certification results. \\
		Process phase where evaluation activities are undertaken.\\
		Evaluator access to intellectual property and artifacts. \\
		Role of architectural decisions and implementation choices. \\
		Process indicators vs. examination of development artifacts. \\
		Reusability of evidence from prior evaluations.\\
		Diversity of kinds of evidence to support judgments.\\
		Up-front investment and ongoing cost.\\
		Benefits to cost, schedule, and risk management. \\
		Enhancements to engineer productivity. \\
		Composability of certification artifacts and results.\\
		Support for certification of ecosystems (libraries, frameworks).\\
		Skill requirements for developers and evaluators. \\ \bottomrule
	\end{tabular}
	\label{tab:setofCriteria} 
\end{table}


%
Second, we analyzed the latest editions of the Common Criteria and DO-178C standards (a total of over 1300 pages) \cite{commonCriteria:CC, commonCriteria:CEM, do178c}. The process of reviewing the documents involved understanding the standards and mapping relevant parts from the documents to the \concept{} we initially identified in our set. For example, regarding the ``Reliability and validity of certification results'' \concept, we identified that the language used by Common Criteria \cite{commonCriteria:CC} to define how evaluators should assess products design leaves much room for interpretation: \QUOTE{The evaluator shall determine that the design is an accurate and complete instantiation of all security functional requirements.} This specific example motivated us to ask questions about practices around inter-rater reliability during our interviews.

Third, we used the initial set of \concept{} and the technically-informed analysis of standards' documentation to guide the design of our semi-structured interviews. We designed our interview questions to allow differences around the \concept{} between the two standards to emerge during the interviews. During an interview, we start with an open question such as `\emph{How long does a typical certification process take?}' and proceed from there based on the experiences of each interviewee. We shared interview guides across standards, but designed separate interview guides for developers and evaluators (see Appendix \ref{apdx:analysisStdDocuments} for examples of questions asked to evaluators). We conducted interviews with \NRParticipants{} participants (see Section \ref{sec:participantRecruitment}) experienced with the standards. We refined our interview questions after each interview and more substantially after a thorough data analysis conducted after nine interviews. We aimed to identify topics that we had not covered in the interviews or for which we received evasive, vague, or conflicting answers from participants. Interleaving interviews and data analysis was essential to grounding the differences across the two standards in practice. 

Finally, in addition to interviews, we read and coded white papers, experience reports, and opinion papers by consultants and subject-matter experts (see Section \ref{sec:paperSelection} for details). We used the surveyed papers as additional interview data by extracting interesting comments, insights, and quotes about issues with certification efforts that map back to our initial set of \concept{}. 

\subsection{Data Analysis}

Our data analysis was based on Qualitative Content Analysis \cite{schreier:2012QCAbook} and we used coding both as reductive and conceptual device. That is, we use it to reduce our text-based data from interview transcripts and from surveyed papers, but also to draw connections between our data and the set of \concept{} relevant for standard comparison. The goal was to build a coding frame to structure data obtained from the cross-validation of our set of \concept{} with interview transcripts and surveyed papers. All interviews and papers were coded by at least two authors and all new or conflicting codes that emerged were discussed and merged when necessary.

\begin{figure}[t]
	\centering
	\includegraphics[width=\linewidth]{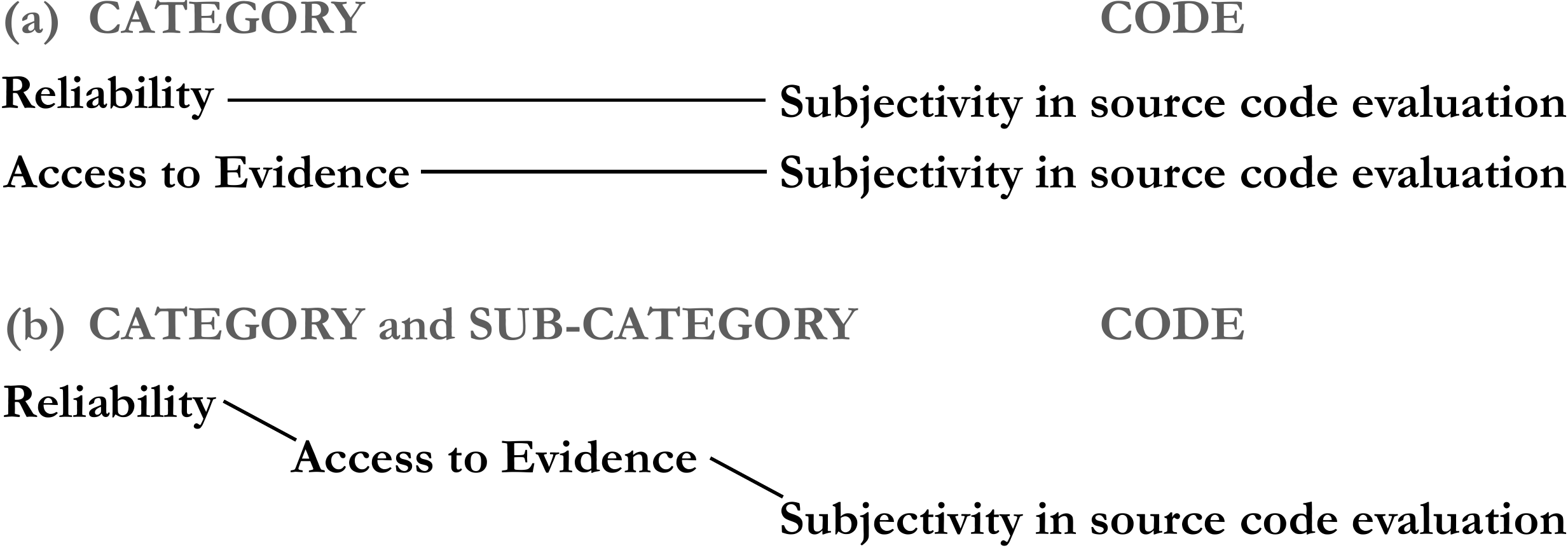}
	\caption{Example of code and categories obtained from iterative content analysis of textual data, which includes interview transcripts and literature papers (a) and a slice of the final coding structure after codes and categories have been merged and organized hierarchically (b).}
	\label{fig:coding-frame}
\end{figure}

To explain in more detail how we built the coding frame obtained from the analysis of interviews' and surveyed papers' raw data, we will use an example of the design quality \textbf{Reliability}. To build our concept-driven coding frame \cite[p.~84]{schreier:2012QCAbook}, we first used our initial set of \concept{} as code categories, as described in Table \ref{tab:setofCriteria}. The categories from the initial coding frame were unstructured, directly reflecting the set of individual \concept{} we identified during the discussion phase. Next, we refined these categories iteratively using the raw data we collected from the standards documents, interview transcripts, and papers. We constantly revised our coding frame while incorporating new interview data to our analysis, generating new categories and sub-categories, but also merging old ones. We eventually applied a hierarchy to the coding frame to enable better and deeper comparisons. For example, this allowed us to classify the top-level categories of our coding frame into design dimensions and qualities, which are the central points of discussion in this paper. Figure \ref{fig:coding-frame}(a) illustrates how the raw data is abstracted to codes and (b) shows a slice of the hierarchical coding structure showing \textbf{Reliability} as the top-level category. \\

\myparagraph{Raw data:} \QUOTE{So, we really said, ``How do we add value'' How do we gain assurance so that we can truly mutually recognize these evaluation activities? And that's when \textbf{we decided that review of source code was risky to vendors, was time-consuming and expensive to evaluation facilities, and it really did not give us the confidence that we needed because we had no sense of consistency}.} \\

\myparagraph{Code:} \textit{Subjectivity in source code evaluation};\\
\myparagraph{Categories:} \textbf{Reliability}, \textbf{Access to Evidence}. \\
From this code, which summarizes the snippet in bold font extracted from the quote, we inferred that the discussion is about reliability among evaluators (\textbf{Reliability}), but about the use of\emph{source code}, which relates to the kinds of evidence available to evaluators and to how they are manipulated by evaluators (\textbf{Access to Evidence}).

\subsection{Participant Recruitment}
\label{sec:participantRecruitment}

We interviewed \NRParticipants~ experts with experience in either certification standard in different roles and from different backgrounds (i.e., not all from the same organization). Given the sensitive and political nature of the topic, recruiting participants was significantly more challenging than in our prior interview studies on software engineering topics, but still within the range of what is considered acceptable in high-quality research \cite[p.~189]{creswell:2008:researchDesign, mason:2012:howManyInterviewsIsEnough}. We reached out to additional potential participants using public listings of testing laboratories, contacting companies with certified products, contacting people mentioned in publicly available reports, and used our professional connections to find participants. We did only marginally rely on our professional network to conduct the study (5 out of \NRParticipants~ participants).

As part of the recruiting e-mail, we explained the purpose of the study, asked them about their experience with the standards, and asked to schedule phone or Skype calls with them. After agreeing to participate, we conducted interviews, each lasting 30--40 minutes. 
With participants' consent, all interviews were recorded and later transcribed. Table \ref{tab:participants} lists our participants, their roles in the certification process, and standard expertise. Participants were from 14 different organizations, from seven different countries.

\begin{table}[t]
	\centering
	\small
	\caption{\small Interviewees and their role in the certification process.}
	\begin{tabular}{@{\hspace{.2cm}}l l l@{\hspace{.2cm}}} \toprule
		Codes & Role & Expertise \\ 	\midrule
		CC1, CC6, CC7 & Developer & CC \\ 
		CC3, CC4, CC5, CC9, CC10 & Evaluator & CC\\ 
		CC2 & Meta-evaluator & CC \\ 
		CC8 & Policymaker & CC\\
		DO1, DO3, DO4, DO6, DO8 & Developer & DO-178C \\ 
		DO2, DO5 & Evaluator & DO-178C \\ 
		DO7 & Meta-evaluator & DO-178C\\ \bottomrule
	\end{tabular}
	\label{tab:participants} 
\end{table}

\begin{table}[t]
	\centering
	\small
	\caption{\small Surveyed papers that reflect on either social, technical, or economic aspects of software certification.}
	\begin{tabular}{ll} \toprule 
		Standard & Citation \\ 	\midrule
		Common Criteria & \cite{hearn:2004, maibaum:2008, lipner:1991, keblawi:2006, lipner:2015birthAndDeathOrangeBook, lipner:1999, shankar:2004}\\ 
		DO\-178C & \cite{thomas:2007, dodd:2012, rushby:2011, heimdahl:2007, moy:2013testingFormalVerif, hatcliff:2014}\\ \bottomrule
	\end{tabular}
	\label{tab:literaturesurvey} 
\end{table}

\subsection{Literature Survey}
\label{sec:paperSelection}
To complement our interviews, we conducted a literature survey and selected white papers, experience reports, and opinion papers 
that discuss practices, challenges, and solutions to problems related to software certification. Our paper selection is targeted, since we focus on papers that \emph{reflect} on either social, technical, or economic aspects of software certification. In total, we selected \NRPapers~papers, seven papers discussing Common Criteria and six papers discussing DO-178C (see Table \ref{tab:literaturesurvey}). 

We searched the web (Google Scholar) and digital libraries (IEEE Xplore, ACM), using ``\emph{Common Criteria}'' and ``\emph{DO-178C}'' as the main keywords and went through the first 100 results and read each paper's title and abstract, discarding clearly out-of-scope items. That is, we intentionally surveyed papers that \emph{critically reflect} on certification standards and their processes and discarded many papers that:
\begin{itemize} 
  \item only provide an overview (and explain the basics) of each standard,
  \item only lists changes in certification processes overtime, or 
  \item only discuss narrow technical aspects of a particular standard.
\end{itemize}
Furthermore, we used snowball sampling to identify further relevant papers from the reference list of papers as we read them and added two papers suggested by our interviewees. The distinctiveness of the standards' names reduces the risks of missing important papers. After having downloaded each selected paper, we transform them into text and analyze them as interview transcripts.

\subsection{Threats to Validity}
Our study may suffer from threats to validity commonly found in qualitative studies. 

Regarding external validity, one needs to be careful when generalizing beyond the results of analysis of the two standards. The differences between the standards may be limited since the reasons they were selected were the same---their community, their popularity, and the availability of data about them. 

Regarding internal validity, one needs to be careful about potential theory, researcher, and selection biases. Our \emph{initial set of criteria} to compare standards---obtained at early stages of research as a result of informal conversations with stakeholders---guided our entire collection and analysis processes. We, however, have not forced the opinion data obtained in interviews and literature survey to match the initial set of criteria, allowing new criteria to emerge and old criteria to be refined in our analysis. Our theoretical background about certification processes, and more specifically about the two certification efforts, could affect the researcher interpretation during standard document analysis, data collection, and data analysis. When analyzing standard documents with fresh eyes---like we do in this study---we increase the chances of a researcher to perceive differences across cases, but also increase the potential for misunderstandings about the studied topic. Even though we have not observed substantial coding disagreements and have worked together to settle small ones, one needs to be careful when generalizing the reliability of coding results since only a pair of researchers was involved in each transcript analysis.

Our results may be affected by selection bias, since the opinions from who have not agreed to be interviewed---who found the matter too sensitive to speak with researchers--- and from who have not written papers about their experiences may have had different opinions and insights about the certification standards. The distribution of participants is slightly skewed toward Common Criteria. When analyzing the distribution of interviewees roles, the distribution is skewed toward evaluators for Common Criteria and toward developers for DO-178C. This was not intentional and can be possibly explained by the availability of public information about testing laboratories and by our broader personal network of software engineers working on aerospace industry. The qualitative nature of our analysis and results should only be mildly affected by this potential bias.

\section{Certification Quality Aspects}
\label{sec:correctness}

From our combined analysis of interviews, standard documents, and literature, we identified \NRDesignDimensions~ \emph{design dimensions} and \NRDesignQualities~ \emph{design qualities} that affect certification efforts. The set of dimensions and qualities as well as their relationships were derived from the top-level categories of our resulting coding frame, as explained in Section \ref{sec:researchMethod}. In this study, \emph{design dimensions} refer to ranges of alternatives in the design space from which a designer can explicitly make decisions about properties of the software products being certified, the certification processes and the roles vendors and evaluators played in them, and the standards themselves. Alternatively, design qualities refer to ranges of outcomes in a design decision-making process. That is, \emph{design qualities} can be interpreted as independent variables of the design process that uses \emph{design dimensions} as dependent variables. Table \ref{tab:dimensions} describes the ranges for the found \emph{design dimensions}.

The final codes in our analysis included both design dimensions and qualities. We chose to organize and present our results in terms of the relationships between them, using design qualities as indexes. Each following sub-section of the paper discusses a design quality, in which we discuss the studied standards in context and provide grounded data about current practices and trends around the identified \emph{design dimensions}. Table \ref{tab:dimensionsQualities} summarizes the relationship between the identified dimensions and qualities, where rows represent design qualities and the columns represent dimensions. The presence of an \myuparrow{} or \mydownarrow{} in the table indicate a positive or negative interaction between them. For example, in a hypothetical situation where one wants to maximize the reusability of evidence across product evaluations, explicitly choosing to increase evaluation frequency can facilitate it. The \myquestionmark{} indicates that no clear relationship was identified, and further investigation is necessary. The absence of an arrow/discussion between a dimensions and a quality does not imply that an interaction does not exist (neither implies that it exists), but only that the topic was not identified in our interviews or literature for the two studied standards. The following sub-sections contain both nuanced discussions of the results---which include viewpoints collected from \NRPapers~ papers and \NRParticipants~ interviews---and our interpretation of the results. Table \ref{tab:dimensionsQualities} intends only to roughly outline the findings---which is consistent with case study research \cite{flyvbjerg:2006}.

After presenting our results, we discuss implications about the actual development and evolution of certification efforts and explain how stakeholders could benefit from our analytical framework in Section~\ref{sec:discussion}.

		\begin{table}[t]
	\centering
	\small
	\caption{\small Design dimensions and their range of options.}
	\begin{tabular}{@{\hspace{.1cm}}p{2.8cm} p{4.4cm}@{\hspace{.1cm}}} \toprule 
		\textbf{Name} & \textbf{Description} \\ \midrule
		\textbf{Formality} & How formal are the types of evidence supported by the standard; ranges from informal types often produced manually by developers (e.g., documentation, informal diagrams) to testing, to more formal and sound techniques (e.g., proof-carrying code and proof checkers).\\[.4em] 
		\textbf{Automation} & How automated is the process of producing and evaluating evidence; ranges from manual (involving interpretation and subjective judgments) to fully automated (being directly extracted from software artifacts). \\[.4em] 
		\textbf{Composition} & How much support to composition is defined by the standard; ranges from none, to coarse-grained (e.g., OS+DB), to fine-grained (e.g., plugins). \\[.4em] 
		\textbf{Access to Evidence} & How much access to evidence evaluators have; ranges from incomplete (e.g., sample-based or black-box testing) to full access (e.g., source code). \\[.4em] 
		\textbf{Involvement Style} & How early in the development process are evaluators expected to be involved; ranges from early pre-development involvement to after-the-fact. \\[.4em] 
		\textbf{Evaluation Frequency} & How frequent are evaluations performed; ranges from one-time-snapshot evaluations to continuous monitoring where recertification is performed more frequently. \\[.4em] 	
		\textbf{Standard Rigidity} & How much flexibility vendors and evaluators have in creating alternatives to meet standard objectives; ranges from strict standard compliance to flexible compliance where they can innovate with alternative forms of evidence. \\  \bottomrule
	\end{tabular}
	\label{tab:dimensions} 
\end{table}

\begin{table}[t]
	\def\arraystretch{.2} \small
	\caption{\small Design dimensions (columns) and quality aspects (rows) that affect standards and certification efforts.}
	\begin{tabular}{rp{0.001cm}p{0.001cm}p{0.001cm}p{0.001cm}p{0.001cm}p{0.001cm}p{0.001cm}}
		~ & \rotatebox[origin=l]{67}{Formality}          & \rotatebox[origin=l]{67}{Automation} 	   & \rotatebox[origin=l]{67}{Composition} 
		& \rotatebox[origin=l]{67}{Access to Evidence} & \rotatebox[origin=l]{67}{Involvement Style} & \rotatebox[origin=l]{67}{Evaluation Frequency} & \rotatebox[origin=l]{67}{Standard Rigidity} \\\toprule \\
		Validity 				   & \myuparrowt  	& \myuparrowt 	& ~ 			 & \myquestionmarkt 	& ~ 				& ~  					& ~ 						\\
		Skill Requirements         & \mydownarrowt  & ~  			& ~  			 & ~ 					& ~ 				& ~  					& ~ 						\\
		Reliability                & ~  			& ~  			& ~ 			 & \mydownarrowt 		& ~ 				& ~  					& \myuparrowt 				\\
		Timely Certification  	   & \myuparrowt 	& \myuparrowt  	& ~ 			 & ~ 					& \mydownarrowt 	& ~ 					& ~ 						\\ 
		Certification Risks        & ~  			& ~  			& ~ 			 & ~ 					& \mydownarrowt 	& ~  					& ~  						\\
		Costs with Suppliers       & ~  			& ~  			& \mydownarrowt  & ~ 					& ~ 				& ~ 					& ~ 						\\
		Reusability of Evidence    & ~  			& ~  			& ~ 			 & ~ 					& ~ 				& \myuparrowt 			& ~ 						\\
		Return on Investment       & \myuparrowt  	& \myuparrowt  	& ~ 			 & ~ 					& ~ 				& ~ 					& ~ 						\\ 
		Trust Issues         & ~  			& ~ 			& ~ 			 & \myuparrowt 			& ~ 				& ~  					& ~ 						\\
		Innovation     			   & ~  			& ~  			& ~ 			 & ~ 					& ~ 				& ~  					& \mydownarrowt 			\\\\ \bottomrule
	\end{tabular}
	\label{tab:dimensionsQualities} 
\end{table}


\subsection{Validity}
\label{sec:validity}

Validity, that is, whether evaluations reach a sound assurance judgment, is of central importance and may depend 
significantly on the formality of evidence available to evaluators, on how automated the process to evaluate evidence is, and on how complete the evidence is available for evaluators to make that judgments. 

\textbf{Formality \myuparrow} It is broadly recognized that more formal evidence and analyses generate more confidence in judgements validity, since it enables direct traceability between software artifacts and claimed properties \cite{swForDependableSystemsBook}. However, in practice, we found that these are seldom used, even though they are supported by standards (CC4, DO4). For example, the documentation of Common Criteria loosely specifies the form of evidence that developers are supposed to generate, but it mostly includes many forms of \textit{informal evidence} (often produced manually by vendors)\footnote{As opposed to \textit{formal evidence}, usually derived from software artifacts (e.g., proofs generated from proof-carrying code or requirements coverage obtained from code coverage analysis).}, such as textual documentation describing requirements and informal design and architectural models, but also more formal evidence such as testing evidence, both test protocols submitted by developers and independent (black-box) test results conducted by evaluators (CC1, CC3, CC5). A current trend is to reduce the expectations toward textual documentation in favor of testing, decreasing the amount of informal evidence produced (CC3, CC8), and consequently increasing the validity of evaluation results. For DO-178C, testing-based evidence plays a major role in practice \cite{faa:SoftwareAssuranceSurvey}. The focus on testing is to provide evidence for code coverage (DO1, DO2, \cite{rushby:2011}) and traceability between source code and requirements (DO1, DO2, DO3, DO4). The absence of `dead code', obtained either with line or structural coverage, is a highly valued property to support safety claims of higher criticality software products \cite{rushby:2011}. While testing cannot provide sound guarantees and should not be used alone in assurance judgements \cite{swForDependableSystemsBook}, interviewees stated that it is effective in revealing issues in products (DO4, DO6, DO8). Formal methods and model-based approaches, although supported by the standard, are less common in practice, because they are expensive to be applied and until recently could not completely replace testing efforts (DO1, DO2, DO4, \cite{moy:2013testingFormalVerif}). 
\INTERPRETATION{We found that the use of more informal evidence, such as documentation and informal design models, is currently minimized. More formal kinds of evidence and analyses are available in the standards, but there is little incentive to use them since they are expensive to apply. Developers and evaluators seem satisfied with current testing practices.}

\textbf{Automation \myuparrow} Typically, multiple types of evidence are produced for evaluation reasons and evaluators have to simultaneously handle all of them to verify how security or safety requirements are met. We conjecture that automation can support evaluators in these tasks and increase the validity of judgements by mitigating the amount of human error when manually checking multiple evidence. For Common Criteria, participants state it is typical for evaluators to verify how good or complete the documentation is (both documentation on the system's requirements, structure, documentation of process aspects) and to independently test the products' security functionalities in a black-box fashion (CC1, CC2, CC4, CC5). The evaluation process is mainly manual, but one participant highlighted that some tests can be automated as long as they are not artificially creating results (CC4). DO-178C interviewees indicated that evaluators typically witness the execution of tests and check the traceability between requirements and implementation in the code (DO4, DO8). The strong traceability requirements defined in the standard enable the evaluation process to be more automated. As DO8 puts it: \QUOTE{the process must be repeatable and the traceability is an activity that you can redo.} 
\INTERPRETATION{Our analysis shows that traceability requirements facilitate the automation of assurance judgements. Informal evidence is currently increasingly replaced by testing efforts in an attempt to reduce manual review effort.}

\textbf{Access to Evidence \myquestionmark} When evaluating products, evaluators need access to evidence to make sound assurance judgments. We conjecture that access to a more diverse and complete set of evidence increases the validity of judgments. For Common Criteria, despite many opposite claims we heard before this study, multiple participants from the private sector reported being fairly open with access to source code and other artifacts (CC1, CC9, CC10). That is, usually, they provide evaluators with access to any artifact they judge relevant for the evaluation. For DO-178C, \textit{evaluators are seen in a stronger position than vendors}, since, by the mandated process, they must have access to all evidence produced by developers (DO3, DO5, DO6, DO7, DO8). As DO5 puts it: \QUOTE{The certification authority, they're just in a strong position. They can say, `well, unless you provide the evidence, you won't get the certification,' So, the commercial company will make reasonable effort to provide the evidence.} In practice, to follow the mandated process while still protecting their intellectual property, vendors permit access to source code only in small samples (DO4, \cite{dodd:2012}) and at the development site (DO2, DO3, DO4) or forbid access to source code and make black-box testing mechanisms based on program inputs and acceptance criteria available for evaluators. 

\INTERPRETATION{We observed that security vendors were more open about access to evidence than aerospace systems vendors. It is unclear from our study whether partial evidence shared with evaluators affect their judgements validity.}
\subsection{Skill Requirements and Training Costs}
\label{sec:skillReqs}

The professional background, academic training, and experience of both developers and evaluators influence the quality of the
evidence produced, the usefulness of evidence available for judgments, and the thoroughness in which the evaluation process is conducted \cite{voas:1998triangle}. We explored incentives vendors have to hire developers with specific skills to work on products and the average profile expected for evaluators to work on testing laboratories and certification authorities. This lays a foundation for discussing whether practitioners are ready to manipulate the types of evidence defined by standards and how (see Section~\ref{sec:validity}).

Our analysis does not show any particular concern about developers skills beyond current development practices used in industry. To work with Common Criteria, network engineering, informatics, computer science, and software engineering degrees are among the preferred ones for developers (CC4, CC5, CC9). Security certifications, experience as a vendor or in ``hacking or studying programs'' are also highly valued (CC3, CC4, CC5). In line with Common Criteria's original heavy focus on producing informal evidence such as textual documentation (see Section~\ref{sec:validity}), \emph{verbal and written competency} as well as \emph{endurance} were also mentioned as important skills. 

Concerning skills for evaluators, interviewees from both standards indicate that evaluators usually have from five to ten years of experience on average, although there is no experience requirement. Also, they mention that evaluators often receive training for acclimation to the testing laboratories processes, being especially important for newcomers (DO1, DO3, DO5, DO7, CC3, CC10). 

\textbf{Formality \mydownarrow} It is known that manipulation of more formal evidence and analysis requires specific skills or additional training from developers and evaluators \cite{swForDependableSystemsBook}. However, we found no indication that developers are hired primarily based on skills that align with a standard. Instead, it is more common to provide additional training when developers and evaluators need to develop specific skills to produce or manipulate evidence. For example, training is provided to teach DO-178C developers about strict programming practices that are important to meet safety requirements, such as deterministic resource consumption and nonexistence of dead code (DO1, DO7). 
Model-driven approaches and simulation are supported by the DO-178 and developers are usually prepared to use them to reduce testing effort. However, they often need to teach evaluators about these approaches and convince them the safety requirements are sill met (DO4, DO6). Interviewee DO4 stated: \QUOTE{as they are not part of the industry and not involved in any development, it's very difficult to keep on the track of the industry.} 

\INTERPRETATION{We found no concerns about hiring developers or evaluators with specific skills for the current state of the studied certification standards. The formality of testing mechanisms, currently the most common assurance mechanism used in certification, does not require specific skills from practitioners.}

\subsection{Reliability}
\label{sec:reliability}

When standards allow assurance judgments to have a subjective component, it is important to assure that multiple
evaluators reach consistent judgments, especially when they are competing on 
a market (see also Section~\ref{sec:relationshipLabVendor}). 
\emph{Inter-evaluator reliability} especially becomes an issue when scaling the evaluation process to provide
assurance for many products in a market that cannot be assessed by a single testing laboratory. The failure to achieve 
reliability might lead vendors to seek evaluators that facilitate the evaluation process but compromise its quality.
We explored how evaluators, whether from different laboratories or countries, assure that they reach the same outcome 
for a particular evaluation. 

Despite much effort, Common Criteria seems to struggle to achieve reliability among the different private testing laboratories, as indicated by interviewees that acknowledge observable differences between practices in the U.S and Europe (CC4, CC6, \cite{lipner:1999}) The standard
is explicitly designed for mutual recognition of certification results, replacing the quest for mutual recognition of product evaluations ended in 1998 \cite{lipner:1999}. Hearn \cite{hearn:2004} mentions international harmonization and national investments as the main barriers for comparability of evaluations. In fact, some of our interviewees criticized the mutual recognition scheme as not functional (CC4, CC8) and infeasible for political and economic reasons (CC2). It is specially difficult to reach consistency with high-complexity systems and EAL-based evaluations (CC2, CC4, CC7, CC8), indicating a lot of variance in the judgments made by evaluators (CC2). Not even the oversight structure defined by the certification scheme (described in Section~\ref{sec:commonCriteria}) solves the problem. As CC2 states: \QUOTE{Vendors may give an argument, it may be wrong, but if you don't have the design evidence to back up the argument then from an oversight perspective they look the same. You know the arguments are fine where one may be protecting the information and the other may not.} For DO-178C, interviewees paint a much more positive picture of inter-rater reliability. One interviewee reports perceiving evaluators from different countries \QUOTE{as doing the same things} (DO3), as a consequence of the explicit effort of testing laboratories, both from private companies and from public agencies, across the world to be more consistent.

\textbf{Access to Evidence \mydownarrow} We conjecture that by having access implementation artifacts, evaluators can eliminate the potential imprecision that emerge from the level of indirection and subjectivity present in informal evidence, potentially increasing reliability among assurance judgments. From the interviews, we observed that the use of implementation artifacts depends directly on the experience of evaluators (CC2, CC3, CC5, CC9, CC10), but most of them prefer not to use implementation artifacts in their judgements. CC8 explains: \QUOTE{Source code is not providing – source code review does not provide us an adequate return on investment for the risk associated}. CC8 continues: \QUOTE{We decided that review of source code was risky to vendors, was time-consuming and expensive to evaluation facilities, and it really did not give us the confidence that we needed because we had no sense of consistency}. In an effort to improve reliability among evaluators, Protection Profiles---and the test baselines defined by them---are becoming more common for evaluations (CC2, CC4, CC8). CC8 explains: \QUOTE{We really had to kind of come to a place where we looked at really kind of crystallizing what exact test activities did we want conducted as part of these evaluations.} 
\INTERPRETATION{We observed that the direct manipulation or review of source code and other implementation artifacts are sometimes avoided to increase reliability among evaluators, aiming to reduce the amount of what are considered time-consuming and expensive activities to testing laboratories.}\\

\textbf{Standard Rigidity \myuparrow} We argue that by being more specific about how evaluation activities should be conducted and by investing in \emph{harmonization} practices, certification efforts can achieve higher reliability. For DO-178C practitioners, the use of implementation artifacts does not seem to be an issue, since they feel safety properties can be more objectively assessed in the source code (e.g. in terms of code coverage) (DO4, DO7). In addition, participants mentioned many strategies that are used to promote common understanding between evaluators: policies and position papers that define the use of specific technologies, face-to-face meetings and workshops every year among evaluators from different countries, and continuous training are used to ensure everyone has the required competency to perform evaluations (DO4, DO5, DO7). 

\INTERPRETATION{We found that specifying more objectively in the standard how implementation artifacts should be used in an evaluation and consistently promoting harmonization among evaluators have positive effects on reliability.}\\

\subsection{Timely Certification} 
\label{sec:time}

Interviewees of the two standards had different views on the urgency of releasing products and on whether the evaluation process duration is an issue, with Common Criteria developers much more concerned about timely certification than DO-178C stakeholders. 

\textbf{Involvement \myuparrow}. We conjecture that an early involvement of evaluators in the certification process can increase the chance of developing products that are already compliant to a standard, and consequently, reduce certification time. 

For DO-178C, the evaluation time is aligned with the time required to develop complex aviation products, such as airplanes or satellites. The development can take years, usually from three to five depending on their complexity (DO6, DO7), giving vendors enough time to think and plan the evaluation of their products. Because vendors of aviation products do not have a choice of whether to certify their products or not, evaluators are involved in the development and certification from the beginning\footnote{\emph{Stages of Involvement} audits (SOIs) are expected throughout the software life cycle of a project \cite{faaJobAid}.} and vendors know that they cannot rush a product to the market. In contrast, Common Criteria vendors often have finished products, and sometimes sales pending on a certificate, which sometimes gives little flexibility for vendors to re-engineer their products and for evaluators to suggest security fixes at the design level. When vendors have sales pending on a certificate, the average one-year evaluation cycle is less attractive (CC3, CC6). 

\INTERPRETATION{We found that the phase in which evaluators are involved in the development and certification of products generate different expectations around time for evaluation. The less overlap between the development and the certification of a product, the higher are vendors' expectations toward more expedite evaluations.}\\

\textbf{Automation \myuparrow} We argue that automation can support certification efforts in reducing the average time to evaluate products. 
DO-178C certifications can take longer than usual at higher levels of assurance (DO3, DO4, DO6). This is driven largely by the coverage requirements of the standard (DO2, DO5, DO4), \textit{which can be reduced with existent specialized tooling to automate traceability checks} \cite{wolf:2015}. While the safety community that builds aerospace products seem to be more stable regarding the automation of their processes, the security community seeks constant revisions of their standard to increase the use of regression testing in the evaluation process rather than relying on manual documentation review to enable a more expedite evaluation process. 

\INTERPRETATION{We found that tools play an important role in automation when certification processes are more stable.}\\

\textbf{Formality \mydownarrow} We conjecture that informal evidence tend to require more interpretation from evaluators, which can delay the evaluation of products. In fact, among causes for long evaluation cycles observed in Common Criteria certifications, interviewees report \textit{reviewing issues and fixing non-conformities in evidence submitted to evaluation}. For example, it is known that it is difficult to keep security documentation aligned with complex evolving systems requirements and design \cite{keblawi:2006}. One goal behind the recent push toward Protection Profiles in Common Criteria is to reduce certification time from years to months by replacing the extensive amount of documentation required by \emph{domain-specific test requirements} (CC4, CC8). A policymaker, CC8, characterized this in our interview as \QUOTE{we need to do test activities that are going to give vendors a good return on investment.} 

\INTERPRETATION{We observed a trend indicating a reduction in the amount of informal evidence produced during evaluations, aiming at reducing the average evaluation time.}

\subsection{Certification Risks and Evaluation Independence}

When defining standards and certification processes around them, a key design decision is in when to involve evaluators in the design and development process, which can directly affect the risks and independence aspects of evaluations. There is a range of possibilities, from assessing already early designs to assessing only the finished product after the fact.

\textbf{Involvement \myuparrow} We argue that early involvement can have far greater influence in the success of a certification. For example, evaluators can encourage best practices, good designs, and modern tooling, rather than reverse engineer final products. At the same time, it can harm the evaluation independence aspects of certification efforts. The Common Criteria standard documents specify that final certifications happen after the fact, but that some portion of the evaluation or preliminary evaluations can occur already during design and development. In practice, in most cases, the evaluation is conducted after the fact, often when the vendor learns that certification is required to sell on specific markets. At that point, evaluators have none or little leverage to influence in a product's design. The evaluators job is then performing after-the-fact gap analysis to find non-conformances between the specification and implementation of the products (CC3, CC4, CC5, \cite{shankar:2004}) and even producing evidence on vendors' behalf (CC3, CC4). CC3 explains: \QUOTE{The way I like to put it is product vendors are in the business of making product. They're not in the business of writing certification documents. So, as part of Common Criteria there are still some consultancy aspects.} However, there are exceptions and sometimes testing laboratories are able to provide informal guidance to vendors about how to achieve the security requirements before or while the product is developed (CC5), specially for high-budget projects. One of the interviewed evaluators explained that especially ``mature vendors'', who have experience with prior evaluations, start working with certification in mind starting very early in the development process (CC4), reducing certification re-work costs and risks. 
For DO-178C, interviewees confirmed what the standard only vaguely expects: audits and physical visits by evaluators are common starting early in the development of products (DO3, DO4, DO5, DO6, DO7). The frequency of audits increases as the criticality of products increases, with an average of three physical visits per year (DO6); High-criticality products receive the most attention from evaluators (DO4, DO5). Interviewee DO5 explains: \QUOTE{When we have a project that we really want to be successful, then we get involved very early in the project and look very carefully at the early stage, having the requirements and design reviewed, test readiness reports, review boards, qualification, and all that at great level of detail.} 

\INTERPRETATION{We found that DO-178C regulations and policies require evaluators to be involved early in the development of products. For Common Criteria, however, it is often the case that vendors are not mature enough or are not aware of Common Criteria until they need to sell their products, which increases certification risks.}

\subsection{Costs with Suppliers}
\label{sec:composition}
Another sore topic in our interviews was the costs involved in reusing of libraries, frameworks, and other artifacts from third-parties. A common strategy used in any software development, which includes certified products, is to reuse artifacts to reduce development time and cost. The support for composition mechanisms by certification standards is frequently discussed among practitioners and policymakers.

\textbf{Composition \myuparrow} It is well-accepted that reusing composable artifacts can reduce development time and cost. In certification, however, fine-grained composition actually increase costs and negotiation overhead with suppliers. Although challenging, the composition of functionality is common in practice and supported by both standards at a coarse level of granularity---e.g., composition of application software with a certified database system and a certified operating system (CC2, CC5, DO7, CC6, \cite{hatcliff:2014}). As CC5 puts it: \QUOTE{the developer is required to make a statement of compatibility and from it can be pretty much understood how the two products are connected, what functionalities the composite product uses from the platform.} In practice, composite evaluations have an expiration date to force periodic reevaluation of interactions among products (CC5). Fine-grained composition with third-party dependencies (libraries, services), though, is seriously challenging. The use of external dependencies is influenced by the size and the criticality of the products to be certified. High assurance components tend do avoid dependencies (CC1, DO1, DO2), whereas dependencies are more common in small and low-assurance products where they are not certified (CC1, DO2). DO2 exemplifies that the entertainment system in an airplane may use an uncertified streaming library. If a commercial component should be reused, a vendor can request safety or security evidence from the third-party supplier, often at a premium (DO3). If that component is not already certified, the vendor may need to provide own evidence for that component as well, increasing the cost of certification (DO7). In these cases, commonly is some negotiation between vendor and supplier, in which usually vendors end up paying suppliers for the costs of certifying their products (DO3). In other cases, for example, when vendors need to use certified libraries (e.g., math or logging), the library can be used after it has been modeled and simulated in a model-based development environment (DO2, \cite{heimdahl:2007}), such as SCADE \cite{boulanger:2015scadeBook}. 

For Common Criteria, policy-makers have an increased interest in having mechanisms to support fine-grained composition (CC8), for example,
to enable a certified crypto module to be integrated into a larger product, such as a browser. There is already a synergy between FIPS 140
(a cryptography standard) and some Protection Profiles that support this kind of composition. However, there is still a lack of compositional
assurance mechanisms to reduce the reliance on vendors to follow effective security engineering principles when composing commercial
off-the-shelf products \cite{lipner:1991, hearn:2004, keblawi:2006}. Current mechanisms are limited to testing known interactions between products (CC2, CC5, CC6, CC7, CC8). 

\INTERPRETATION{We observed that support for platforms, ecosystems, and app markets, where an end user may compose (possibly certified) artifacts from different sources are not in scope of either certification standard and are considered incompatible with current assurance techniques based on testing.}

\subsection{Reusability of Evidence}
\label{sec:reuseEvidence}
A common concern in our interviews were issues regarding the maintenance of certified products over time, 
from small incremental changes to larger changes between releases. Both standards have provisions for recertification, expecting
developers and evaluators to perform an impact analysis, which includes adequately documenting the changes in the product being maintained. 

The degree of reuse of evidence is a result of negotiation between vendors and evaluators and it usually involves arguments about the \textit{extent of the changes} and the \textit{time passed since the product was evaluated}. Again, trust between vendors and evaluators plays an important role (see also Section~\ref{sec:relationshipLabVendor}). For minor changes in certified products, participants report \QUOTE{reusing as much evidence as possible} (CC1, DO3, CC9, CC10). Previously unmodified parts of the product are not reevaluated (DO4, DO5, CC5, DO7) and regression tests play an important role supporting the identification of artifacts impacted by changes (DO1, CC4, CC7). For major changes, a complete recertification is expected with each release (CC1, CC6, CC7). In Common Criteria, significant changes in a Protection Profile can require a complete recertification. However, grace periods exists with which a certification can be maintained despite smaller changes: \QUOTE{products can be maintained for two years before they have to be re-evaluated} (CC4). However, the situation is, at least, alarming when vulnerabilities are found in certified security products: \QUOTE{If your product uses an affected version of OpenSSL and it has already been certified or validated, you are not required to recertify it} \cite{corsecHeartbleedBlogPost}. Beyond the reuse of concrete evidence, the \textit{reuse of knowledge} across evaluations emerged as a common topic in our interviews.

\textbf{Evaluation Frequency \myuparrow} We conjecture that by pushing evaluation frequency to what would be a continuous evaluation processes, rather than a one-time discrete one, could minimize the differences between versions of a certified product and potentially increase the reuse of evidence. Although provisions for recertification exists in both standards, practitioners still complain about the costs and duration of recertification (CC6, CC8, DO7, \cite{lipner:1991}). For example, certification is considered to have minimal value if the evaluation process takes more time than product life cycle. The current certification practices are incompatible with frequent releases that are increasingly adopted with the DevOps movement in large parts of the software industry. CC7 explains: \QUOTE{(rapid recertification) is going to matter even more because everything is going faster. My favorite example is containers. If you’re doing containers correctly, you’re talking about doing a build daily or even hourly in some cases}. 

\INTERPRETATION{We found that vendors are concerned about one-time-snapshot certifications, which usually take a long time and provide little flexibility to products evolution. Consequently, we expect one-time-snapshot certifications to be prohibitive in scenarios that expect frequent product releases or major changes in products overtime.}

\subsection{Return on Investment}
\label{sec:returnOnInvestment}

With certification processes being perceived as costly and slow, there is a question whether the perceived benefits justify the costs---whether certification has a positive return on investment. The literature reported that vendors do not believe that certification improves the
quality of evaluated products or processes \cite{hearn:2004, keblawi:2006, maibaum:2008, lipner:2015birthAndDeathOrangeBook}. Some of our interviewees report hearing the same complaints from vendors (CC3, CC6). At the same time, interviewees actually often have something positive to say about certification. For example, as direct benefits participants mentioned an increased understanding of the products by developers (CC1, DO1) and a noticeable quality improvement of evaluated products (CC3, CC6, CC7).

\textbf{Formality \myuparrow} We conjecture that more formal evidence would be considerably more expensive to produce and not provide good return on investment. However, DO-178C interviewees explained a positive feedback cycle in which certification costs pushed developers toward increased use of model-based approaches to verifying requirements before implementation, which also reduced rework of issues propagated to code and tests (DO3, DO4). There is also an incentive for developers to adopt such approaches to reduce testing efforts, though evaluators are not necessarily technically prepared to evaluate products using this kind of technology (see Section~\ref{sec:skillReqs}). 

\INTERPRETATION{We found that some types of more formal evidence, such as models that can be simulated, can enable greater return on investment for vendors. However, evaluators are not necessarily technically apt to use them.}\\

\textbf{Automation \myuparrow} By increasing the repeatability of processes and reducing development and evaluation costs, we conjecture automation can enable greater return on investment. For Common Criteria, some interviewees reported a significant value from the evaluation process, since test suites were expanded to satisfy more rigorous testing requirements \cite{shankar:2004} and many bugs were fixed during regression testing activities. (CC6, CC7). For example, CC6 mentioned how the Linux random number generator benefited greatly from going through certification and being extensively tested and reviewed in the evaluation process. 

\INTERPRETATION{We found that increased repeatability, even if not using formal evidence, can enable greater return on investment for both vendors and evaluators that can re-run tests across different versions of certified products.}

\subsection{Trust Issues}
\label{sec:relationshipLabVendor}

Access to intellectual property (e.g., details of a proprietary encryption algorithm embedded in a security product), can cause real trust issues between vendors and evaluators in the certification process.

\textbf{Access to Evidence \myuparrow} We argue that requiring vendors to share more evidence, specially implementation artifacts, can generate trust issues between vendors and evaluators. For example, vendors might fear leakage of proprietary information to competitors by evaluators. However, we found that vendors can be fairly open about access to intellectual property or use other mechanisms to protect it, as discussed in Section~\ref{sec:validity}. Also, evaluators often prefer not to use implementation artifacts (see Section~\ref{sec:reliability}, which potentially minimizes trust issues. One participant mentioned the use of implementation artifacts to protect testing laboratories from liabilities. CC5 explain: \QUOTE{we keep a copy of vendors' encrypted source code to protect us in case of court disputes.} Most interviewees emphasize the importance of a trust relationship between vendors and evaluators that forms over time. 

\INTERPRETATION{We observed that vendors tend to be open about implementation artifacts, but can rely on protection mechanisms when trust has not been developed between vendors and evaluators and there are concerns about leakages of intellectual property.}

\subsection{Innovation}
\label{sec:innovation}

Innovation plays an important role in certification efforts, since software represents both the means and end of certification. Software systems are frequently evolving or being replaced by more modern versions that contain new technology. Hence, it is expected that standards, process, and tools also evolve to correspond to systems being developed. There is also an interesting synergy between innovation and skills requirements. Developers and evaluators are resistant to change their practices, which can be an obstacle in evolving the standard (DO2, DO4, DO7). As discussed in Section~\ref{sec:skillReqs}, DO-178C evaluators often have to be taught about more formal evidence and convinced by developers that the same safety guarantees hold with them. This limits the adoption of formal methods, model-driven approaches, and simulation in practice, since they expect specific technical skills from evaluators.

\textbf{Standard Rigidity \mydownarrow} We conjecture that the lack of flexibility in a standard can hinder innovation and lead to its obsolescence (CC1, \cite{keblawi:2006}). In the past, \QUOTE{federal agencies have abandoned monolithic standards, because they lacked flexibility and led to expensive and complex acquisitions. Standards can prevent the use of highly desirable but nonconforming technologies and COTS products \cite{keblawi:2006}.} Interviewee CC6 warns: \QUOTE{I think that is the danger is that the certification efforts don't keep up with technology everyone's gonna get waivers and no one's gonna care}. In an attempt to mitigate this issue, standard revisions and policies are released to ensure that the latest technology gets out to end customers (CC8, DO7). 

\INTERPRETATION{We found that technology adoption is slow and that evaluators have no or little incentive to maintain consistency with technology used by vendors.}

\section{Related Work}
Relevant work about certification standards is incorporated into our literature survey (see Section~\ref{sec:paperSelection}), discussed as part of our results (see Section~\ref{sec:correctness}), and summarized in this section. The literature underpinning our research design is described in sections \ref{sec:researchDesign} and \ref{sec:relatedWorkDesignSpaces}.
\subsection{Discussion of Common Criteria}
\label{sec:relatedWorkCC}
Several authors have critiqued Common Criteria since its conception, which provides context to why it has been and still is revised. Hearn \cite{hearn:2004} puts Common Criteria in question, sharing the pessimistic opinions of buyers and sellers about the perceived benefits of the Common Criteria certification. He suggests improvements around composition and pre-development certification to better address vendors' needs and to enable products to be re-engineered if necessary. Keblawi and Sullivan \cite{keblawi:2006} provide an interesting perspective of Common Criteria being applied to aircraft traffic control systems. The paper also identifies a need for better security requirements at the system level, including composition of certified products. Lipner has long been involved in the community of security standards in the United States \cite{lipner:1991, lipner:1999, lipner:2015birthAndDeathOrangeBook}. More recently, Lipner \cite{lipner:2015birthAndDeathOrangeBook} presents an overview of the story of Common Criteria (and the old \emph{Orange Book}), providing context for still current issues and aspirations for Common Criteria certification. For example, he highlighted the difficulty in achieving reliability among evaluations and frequently observed misalignment between documentation and actual product implementation. 
As examples of papers about Common Criteria that we excluded from the literature survey, we mention the work from Barabanov et al. \cite{barabanov:2014:russianCC}, Kaluvuri et al. \cite{kaluvuri:2013:webServicesCC}, and Kang et al. \cite{commonCriteriaForSmartTVs:2017}, which discuss Common Criteria being applied by a different scheme/country and technical aspects around the evaluation of Smart TVs and Web Services, respectively. In addition, Shankar et al. \cite{shankar:2004} report their experience in evaluating Linux using Common Criteria. The paper mentions both challenges and positives aspects in conducting this process in open source. The frequent mismatches between code and design documentation is a barrier, but source code availability and promptness of the community in face of new vulnerabilities make evaluations of operating systems easier than in commercial settings.

\subsection{Discussion of DO-178C} 
\label{sec:relatedWorkDO178C}
Stakeholders in the aerospace domain seem much more confident in the DO-178C, but still often critique aspects of the standard, such as the exaggerated focus on hardware-level assurance, the subjectivity in certification audits, and the lack of automation in specific areas of the process. In particular, Thomas \cite{thomas:2007} reacts to a National Research Council report and calls attention to the infeasibility to provide the same level of assurance for software systems as it is usually done with hardware. He suggests improvements for DO-178C in three areas: requiring explicit claims about software properties, providing evidence to support claims at a practical and useful level (as also mentioned by Rushby \cite{rushby:2011}), and determining specific expertise to help developers/vendors in creating useful evidence for claims. Dodd et al. \cite{dodd:2012} discusses a set of issues regarding certification audits of airborne software, which includes subjectivity in certification processes and lack of predictability of the certification process. To mitigate these issues, they propose an statistical method for supporting certification audits based on continuous data analysis of projects' lifecycle to trace potential deviations overtime. Moy et al. \cite{moy:2013testingFormalVerif} report their successful experience using formal methods to replace testing efforts in a production-like environment at Dassault-Aviation and Airbus, showing that it can be practical and faster with the support of more automated tools.
More recently, Hatcliff and colleagues \cite{hatcliff:2014} identified current socio-technical challenges in developing and certifying safety-critical software-systems, after revisiting topics discussed earlier by Heimdahl \cite{heimdahl:2007}. New challenges are mentioned, but old challenges, such as requirements validity, composition, role of tools in certification, automation, and training for developers (often not prepared to produce safety evidence) are still pain points for certification. 
We also found other works that discuss DO-178C, but are mostly descriptive and focus on technical details. Gigante and Pascarella \cite{gigante:2012:formalMethodsDO178C} and Ulrich and Allen \cite{ulrich:2016:verificationTechniquesDO178C} are explain formal methods and verification techniques would fit the DO-178C certification process life-cycle, respectively. Hilderman \cite{hilderman:2014:DO178CBenefitsCosts} describes the major costs and benefits from DO-178C certification.

\subsection{Other Certification Standards} 
\label{sec:relatedWorkOtherStd}
As mentioned in Section~\ref{sec:certifyingSoftSystems}, there are many certification efforts besides the two investigated in this paper, which includes NIST RMF 800 Series \cite{nistRMF800} for information security, IEC 62304 \cite{isoIEC62304} for software medical devices, IEC 60880 \cite{IEC60880} for nuclear software systems, IEC 15504 \cite{isoIEC15504} for software processes, IEC 25000 \cite{isoIEC25000} for software products functional quality, and many others defined by organizations such as ISO/IEC. Rodriguez et al. \cite{rodriguez:2015:hardLookSoftQuality, rodriguez:2016:evalSoftProductFunctionalSuitability} report their experience in using IEC 25000 to evaluate the functional suitability of a web application for managing human resources in two opportunities: after and before some functional changes. The results shows a consistent level of compliance between functional elements evaluated and number of product requirements implemented. These standards could also be compared using a similar process and the same \emph{initial set of criteria} we used to compare Common Criteria and DO-178C (see Table \ref{tab:setofCriteria}), but the investigation of these standards are not in the scope of this paper. 

\subsection{Standards Comparison}
\label{sec:relatedWorkStdComparison}
There are also papers that perform standards comparison, but much more focused and fine-grained. Lahtinen et al. compared both IEC 60880 and IEC 61508 standards, aiming at identifying differences in strictness and scope between the ``shall'' requirements as defined by the two standards. The results indicate that such fine-grained comparison is difficult, but they could still identify that even though the two standards are fairly similar, they complement each other and should be used together to achieve higher safety assurance. Hawkins et al. \cite{hawkins:2013:assuranceCasesDO178C} investigated differences in certifying a wheel braking system when using a prescriptive approach (DO-178C) and when using assurance cases. The authors suggest linking assurance cases and DO-178C processes to increase safety assurance, since using assurance cases alone might be challenging, especially for vendors without experience in developing cases. Our study complements these papers by contrasting two distinct standards, which contributes to a broader discussion about certification efforts across communities. We are not aware of any other papers that compare standards across communities, focusing on contrasting cases to confront community-specific culture. 	

\subsection{Design Spaces}
\label{sec:relatedWorkDesignSpaces}
Design spaces have been widely studied by the design community, where exploring alternatives more systematically is considered good practice \cite{shaw:2012, woodbury:2006:DSWhiter, maclean:1991:DSElementsAnalysis}. Designers are often described as explorers that use their abilities to assess design options and make informed decisions based on the trade-offs that these options entail. With our study, we attempt to provide an initial contribution to building a design space for software certification standards, so that stakeholders can also make informed decisions when creating or updating standards, empowering them to reason more systematically about the options in the space, similar to how many other areas have done in the past with architectural styles \cite{shawClements:1997}, computer input devices \cite{card:1990:DSInputDevices, kern:2009:DSAutomotiveUI}, wireless communication \cite{romer:2004:DSWirelessComm}, 3D architectures \cite{xie:2006:DS3DArch}, and self-adaptive systems \cite{brun:2013:DSSelfAdaptSys}.

\section{Reasoning About Certification Efforts Design}

Our grounded analysis of both literature and interviews with stakeholders in two certification standards can provide initial unbiased guidance to stakeholders, including practitioners, testing laboratories, policymakers, academics, government agencies, or anyone in a position to influence decisions about certification standards and processes, regarding the many complex choices and implications involved in designing certification efforts. 

As already discussed in Section \ref{sec:correctness}, our paper does not propose to generate a complete bounded set of design dimensions and qualities that affect certification efforts, but to contribute the foundation toward a theory that can explain the relationships between them. In this specific case, a resulting theory should resemble the contents of Table \ref{tab:dimensionsQualities}, which describes positive and negative interactions between design dimensions and qualities.

Even in its initial development stage, we argue a theory like this can be useful to support reasoning of the design decisions that can influence the result of certification efforts. The following outlined steps (and the hypothetical scenario described) exemplify how one could
use a theory in concrete scenarios while creating or revising a certification standard:

\begin{enumerate}
	\item select a quality of interest;\\
	 \textit{If a member of certification standard committee decides vendors (developers) should have freedom to innovate with new tools and assurance techniques (for example, to ensure that compliance processes do not become obsolete), they would go to Table \ref{tab:dimensionsQualities} and lookup the \textbf{Innovation} quality row;\\}
	\item check which dimensions affect it;\\
	 \textit{Then, our results would indicate a relationship between \textbf{Innovation} and \textbf{Standard Rigidity}. In this case, in order to increase the opportunity for innovation, they would have to make the standard more flexible in terms of compliance processes, tools, or technology;\\}
	\item for each dimension that affects the selected quality (selected in step 1), reason about how other qualities are affected;\\
		 \textit{In our scenario, the results would indicate that adjusting \textbf{Standard Rigidity} (for example, by making a standard more flexible) could indirectly lead to negative implications on \textbf{Reliability}. By knowing the consequences of making a certification standard less or more rigid, one can reason about solutions that minimize reliability issues while maintaining the standard evolving along new technological trends.}
\end{enumerate}

\vspace{1em}
The outlined steps (and the hypothetical scenario described) shows how a theory---even in its early stages of development---can be used to reason about the trade-offs involved in designing certification. While simplistic, the described scenario highlights the potential of the theory to highlight direct and indirect relationships between design dimensions and qualities. Finally, we call for attention that stakeholders should not only use the summarized results in Table \ref{tab:dimensionsQualities} to make decisions, but also consider text containing the nuanced discussions of the results and decide for themselves, considering the context they are inserted, whether the theory applies.

\section{Conclusion}
\label{sec:discussion}
As our results shows, we have identified many dimensions and qualities that can affect the design of certification efforts.
The options available in each design dimension have distinct implications and can directly influence the technical, social, and economic qualities of certification. More investigation is necessary to elicit other relevant design dimensions, disambiguate relationships between them, and confront our preliminary results. However, our grounded analysis of both literature and interviews with stakeholders in two certification standards can provide initial guidance to stakeholders about the many complex choices and consequences involved in designing certification efforts. 

\myparagraph{Implications.} More generally, observations about one particular standard can be used by stakeholders in another standard to rethink about their own certification processes, techniques, and practices. Also, observations from individuals that play one particular role in the certification process can provide a distinct perspective to stakeholders in different roles, increasing their awareness about issues and enabling them to think about alternative incentives to address potential issues in their certification processes. Specifically, \begin{inparaenum}[(1)]
	\item developers and vendors of one particular standard can now have a distinct perspective of techniques used in another standard, which enables them to reason about the applicability of the techniques in their context and about how it could affect their certification processes and results quality (e.g., DO-178C requirements/code coverage could be practical for Common Criteria vendors to mitigate the subjectivity in their source code evaluation),
	\item evaluators and meta-evaluators can now understand how developers and vendors in a different domain react to standard compliance requirements and what are their practices when producing evidence, which can enable evaluators to develop policies that can refine standards to prevent misuses or abuses of techniques to evade these compliance requirements,
	\item policy-makers can directly use our results to understand trade-offs in certification and make decisions that affect emergent certification initiatives for IoT ecosystems, self-driving cars, wearable medical devices, and others.
\end{inparaenum}

New standards for self-driving cars, for example, are still being discussed. The U.S. Department of Transportation recently released guidelines to self-driving cars and highlighted the importance \emph{``to offer solutions that the industry can implement''} \cite{nhtsaFederalAutomatedVehicles}. The same document describes their expectations toward software updates: \emph{``manufacturers will likely provide software updates for motor vehicles well after they are manufactured and certified.''}, suggesting the need for fast and automated re-certification. Our results show that the studied standards are not ready to address this need, since cars often have a significant shorter development cycle when compared to airplanes and re-certification procedures still heavily rely on manual change impact analysis techniques. We hope that our analytical framework can provide a starting point into such a discussion and collaborate with initiatives of certifying software systems across recent domains such as IoT ecosystems, self-driving cars, wearable medical devices, and others constantly arising in our society.

\ifCLASSOPTIONcaptionsoff
  \newpage
\fi



%
\bibliographystyle{IEEEtran}
\bibliography{sigproc} 

%

%
%
%




\end{document}